\documentclass[super]{rsc-layout}

\usepackage{amsmath,amssymb,graphicx,textcomp}
\usepackage{hyperref}
\usepackage{cleveref}

\usepackage{verbatim}

\usepackage{times}
\usepackage[usenames,dvipsnames]{xcolor}

\usepackage{bm}
\renewcommand{\vec}{\bm}
\newcommand\diff{\mathrm{d}}

\makeatletter
\newcommand\dash{\penalty\@M-\hskip\z@skip}
\makeatother

\begin{document}

\title{Dynamic arrest in model porous media --- intermediate scattering functions }
\author{%
Markus Spanner\footnotemark[1],
Simon K. Schnyder\footnotemark[2],
Felix H\"of{}ling\footnotemark[3],
Thomas Voigtmann\footnotemark[4], and
Thomas Franosch\footnotemark[1]%
{\renewcommand\thefootnote{\fnsymbol{footnote}}%
\footnotemark[2]}%
}

\abstract{Heterogeneous media constitute random disordered environments where transport is drastically hindered. Employing extensive molecular dynamics simulations, we investigate the spatio-temporal dynamics of tracer particles in the Lorentz model in the vicinity of the localization transition. There transport becomes anomalous and non-gaussian due to the presence of self-similar spatial structures, and dynamic scaling behavior is anticipated.  The interplay of different time and length scales is revealed by the intermediate scattering functions, which are sensitive both to the underlying spatial fractal as well as the anomalous transport. We compare our numerical results in the transition regime to a mode-coupling approach, and find that certain aspects are surprisingly well predicted.
 }


\maketitle

\bgroup
\renewcommand\thefootnote{\alph{footnote}}
\footnotetext[1]{%
Institut f\"ur Theoretische Physik, Friedrich-Alexander-Universit\"at Erlangen-N\"urnberg, Staudtstra{\ss}e~7, 91058, Erlangen, Germany}
\footnotetext[2]{%
Institut f\"ur Theoretische Physik II: Weiche Materie,
Heinrich-Heine-Universit\"at D\"usseldorf, Universit\"atsstra\ss{}e 1,
40225 D\"usseldorf, Germany}
\footnotetext[3]{%
Max-Planck-Institut f\"ur Intelligente Systeme, Heisenbergstra{\ss}e 3, 70569 Stuttgart
and Institut f\"ur Theoretische Physik IV, Universit\"at Stuttgart, Pfaffenwaldring 57, 70569 Stuttgart, Germany}
\footnotetext[4]{%
Institut f\"ur Materialphysik im Weltraum, Deutsches Zentrum f\"ur Luft- und
Raumfahrt (DLR), 51170 K\"oln, Germany, and Zukunftskolleg, Fachbereich
Physik, Universit\"at Konstanz, 78457 Konstanz, Germany
}
\egroup

\footnotetext[2]{Correspondence: franosch@physik.uni-erlangen.de}

\section{Introduction}
One of the most striking phenomena of densely packed  matter is the rapid slowing down of dynamics upon cooling or compression as the glass transition is reached. Certainly, the most prominent aspect is the increase of viscosity by orders of magnitude~\cite{Angell:1995} upon changing the temperature within a few degrees or similarly the density by a few percent. At the same time the diffusion of the particles is strongly suppressed suggesting that motion is strongly hindered until essentially a structural arrest occurs. More insight on the evolution of the glass transition is obtained by focusing at time- or frequency-dependent quantities that encode the dynamic processes and the system's response on different temporal scales~\cite{Goetze:1992,Goetze:1999} as can be nicely achieved in colloidal systems~\cite{Pusey:1986}.
 The next step is to probe the sample also at different length scales to unravel the mechanism that is responsible for the drastic slowing down of dynamics. A series of non-trivial phenomena such as the stretching of time-dependent correlation functions, the time-temperature superposition principle, the emergence of a  fast $\beta$-process, where a factorization property of the space and time dependence occurs, have been observed in the last 25 years~\cite{Goetze:1992,Goetze:1999}. A coherent theoretical framework has been provided in terms of the mode-coupling theory (MCT) of the glass transition developed by G\"otze and collaborators~\cite{Goetze:Complex_Dynamics}. Although many predictions of this microscopic theory have been tested successfully  in experiments and computer simulations, the limitations of the MCT approach are poorly understood.

Recent extensions of MCT~\cite{Krakoviack:2005,Krakoviack:2007,Krakoviack:2009,Krakoviack:2011}
 to liquids confined inside a porous medium predict a rich interplay of the glass-transition singularity and the localization phenomenon in disordered structures, parts of which have been verified already by computer simulations~\cite{Kurzidim:2009,Kurzidim:2010,Kurzidim:2011,Kim:2009,Kim:2010,Kim:2010a,Gallo:2009,Gallo:2011}. There the porous environment serves as a frozen host structure that strongly interacts with the dense liquid filling the void space. This scenario appears to be realized approximately in   ion-conducting sodium silicate melts~\cite{Horbach:2002,Meyer:2004,Voigtmann:2006}, where the silicon-oxygen majority component displays diffusion coefficients that are by orders of magnitude smaller than the ones of  alkali ions. Thus the alkali ions essentially meander through an arrested structure and move along certain preferential diffusion pathways percolating the matrix.  The scenario of a multiple freezing transition has been also discussed for
strongly size-disparate colloidal mixtures of soft spheres~\cite{Moreno:2006,Moreno:2006a,Voigtmann:2009,Voigtmann:2011} or Yukawa particles~\cite{Kikuchi:2007}.

In the limit of very small concentration of the liquid inside the porous structure, the system coincides with the Lorentz model, where a gas of non-interacting tracer particles explores a frozen disordered array of obstacles. There the transport properties depend sensitively on the obstacle density and eventually long-range transport ceases to exist as a certain critical density is reached. Extensive
 computer simulations~\cite{Lorentz_PRL:2006,Lorentz_JCP:2008,Bauer:2010,Franosch:2011,Spanner:2011} confirm that the mean-square displacement increases subdiffusively directly at the transition and scaling behavior holds in the close vicinity of the critical density. A self-consistent
mode-coupling kinetic theory~\cite{Goetze:1981,Goetze:1981c} made a series of predictions of the dynamic behavior for all obstacle densities, in particular, it reproduces the long-time anomalies in the velocity autocorrelation functions~\cite{Ernst:1971a,Lorentz_LTT:2007} and yields an accurate estimate for the first-order correction~\cite{Weijland:1968}. Furthermore the theory predicts a transition to a localized phase and suggests spatio-temporal scaling behavior in its vicinity. The mechanism of the suppression of diffusion is due to repeated correlated scattering with the obstacles confining the tracer to the still accessible void space. Yet, in the immediate vicinity of the transition a new phenomenon becomes important. The excluded volume due to obstacles starts to overlap significantly and builds clusters, eventually leading to self-similar structures for the void space. Hence a geometric percolation transition gives rise to a divergent static length scale which ultimately changes the critical behavior.

The purpose of this paper is to explore at the transition and to characterize transport properties in terms of the intermediate scattering function for particles located on the infinite cluster only. There processes on different length scale can be monitored by adjusting the wavenumber. We elaborate a series of scaling relations at the transition both within a simplified mode-coupling approach~\cite{Schnyder:2011} and dynamic scaling hypothesis~\cite{Kertesz:1983} for the universality class of the critical properties or random resistor networks. We compare both sets of predictions to extensive computer simulations and discuss the scaling properties.

\section{Lorentz model}

\subsection{Model definition}

The Lorentz model is a simple model for transport in porous materials, where a point-like tracer moves through an array of quenched spherical obstacles. Introduced by H.~A. Lorentz~\cite{Lorentz:1905} in 1905 to describe the electron conductance in metals, it is used more generally for transport phenomena in disordered systems, and even serves as a microscopic model for anomalous transport in crowded biological media~\cite{Anomalous_ROPP:2012}.
Obstacles are distributed independently in the sample, hence the statistical properties of the structure are simply characterized by the number
density of obstacles $n=N/V$. Here $V=L^d$ denotes the volume of a hypercubic box, ultimately we are interested in the limit of large system sizes, $N\to \infty$, $V\to \infty$, while keeping $n=\mathrm{const}$. 

The disordered environment is explored by a single structureless particle, equivalent to a gas of non-interacting particles.
We consider a hard-core exclusion of radius $\sigma$ between a single obstacle and the tracer. The interaction distance $\sigma$ can be interpreted as the sum of
tracer and obstacle radius, which include the limiting case of an extended tracer in point-like scatterers,
 as well as a point-like particle meandering in a parcours of possibly overlapping spheres.
Hence the dimensionless control parameter is the reduced obstacle density $n^*=n\sigma^d$, where $d$ is the spatial dimension.  
The particle's motion is confined to the \emph{void space} and one anticipates already that with increasing obstacle density the void space decomposes into a collection of smaller and smaller clusters. Above the percolation transition $n^* > n_c^*$ there is no cluster that spans the entire system and long-range transport ceases to exist.

For the tracer dynamics, we employ Newton's equations of motion with a specular scattering at the obstacles. Then the kinetic energy of the tracer is conserved and only
the direction of the velocity $\vec{v}$ is changed in scattering events, such that the velocity $v=|\vec{v}|$ sets the time scale $t_o=\sigma/v$ of the problem.

\subsection{Scaling theory of the critical dynamics}

At the critical obstacle density $n_c^*$, the infinite cluster becomes self-similar~\cite{Stauffer:Percolation} characterized by a fractal dimension $d_\mathrm{f}$, 
which evaluates approximately to~\cite{Jan:1998} $d_\mathrm{f}=2.53$ in 3 dimensions. 
The dynamics on this fractal is expected to become anomalous since the tracer has to explore a network of ramified structures. The simplest quantity probing the transport properties is the mean-square displacement $\delta r_\infty^2(t):=\bigl\langle \Delta\vec R(t)^2\bigr\rangle_\infty$, where $\langle\cdot\rangle_\infty$ indicates averaging different realizations for particles moving on the infinite cluster only and $\Delta \vec R(t)=\vec R(t)-\vec R(0)$ is the displacement of the tracer after a lag time $t$.

Since the percolating void space is scale-free one anticipates $\delta r^2_\infty(t)\sim t^{2/d_\mathrm{w}}$, where $d_\mathrm{w}$ is known as the walk dimension of the system~\cite{benAvraham:DiffusionInFractals}. For the three-dimensional Lorentz model the value $d_\mathrm{w}=4.81$ differs from the one in random resistor networks on a lattice, since transport is dominated by narrow channels emerging in continuum percolation~\cite{Machta:1985,Lorentz_PRL:2006}. The mean-square displacement is connected to a corresponding time-dependent diffusion coefficient $D_\infty(t):=(1/2d)\diff \delta r^2_\infty(t) / \diff t$, and a velocity auto-correlation function $Z_\infty(t):=(1/2d)\diff^2 \delta r^2_\infty(t) / \diff t^2$. In the Fourier domain, the second moment can be transformed to a frequency-dependent conductivity which displays dispersive transport directly at the critical density~\cite{Spanner:2011}. Few is known beyond the second moment, only recently spatial-temporal properties have been discussed~\cite{Franosch:2011}.

A complete characterization of the statistical properties is given by the \emph{van Hove function}\cite{Hansen:SimpleLiquids} $P_\infty(r,t):=\langle\delta(\Delta \vec R(t)-\vec r)\rangle_\infty$, which constitutes the probability for the particle to have traversed a distance $\vec r$ in lag time $t$. In particular, the mean-square displacement represents just the second moment of $P_\infty(r,t)$.

Self-similarity suggests a scaling behavior\cite{Kertesz:1983,benAvraham:DiffusionInFractals}
\begin{equation} \label{eqscaling_propagator}
 P_\infty(r,t) = r^{-d} {\cal P}_\infty(r t^{-1/d_\text{w}})
\end{equation}
for large distances $r\gg\sigma$ and long times $t\gg t_o$.
Thus one expects typical excursions of the tracer of linear extent $\sim t^{1/d_\mathrm{w}}$, larger distances become increasingly rare, suggesting ${\cal P}_\infty(x\gg 1)$ to decay rapidly.

The behavior for small rescaled distance $x\sim rt^{-1/d_\mathrm{w}}$ is inferred by the following consideration~\cite{FCS_scaling:2011}:
The return probability $\Pi(t,w):=\int_{r\leq w}\diff^d r P_\infty(r,t)$ represents the likelihood that the particle remains or has come back within a distance $w$ from the starting point after a time $t$. Provided that this distance is much smaller than the typical excursions, $w\ll t^{1/d_\mathrm{w}}$, this probability is proportional to the accessible volume $\sim w^{d_\mathrm{f}}$ of the infinite cluster. By the scaling law, \cref{eqscaling_propagator}, this implies ${\cal P}_\infty(x\ll 1)\sim x^{d_\mathrm{f}}$.

The \emph{intermediate scattering function} (ISF)
\begin{equation} \label{eqdef_isf}
  F_\infty(q,t):=\langle \exp\left(\mathrm{i}\vec q \cdot \Delta \vec R(t)\right)\rangle_\infty
\end{equation}
is the characteristic function of the displacement of the tracer $\Delta \vec R(t)$ after a lag time $t$.
This quantity is in principle experimentally accessible
by neutron spin-echo spectroscopy or photon-correlation spectroscopy, where $\hbar\vec q$ is identified with the momentum exchanged between the neutron/photon and the sample~\cite{Hansen:SimpleLiquids}. Observing that the ISF represents merely the Fourier transform of the van Hove function, $F_\infty(q,t)=\int \exp(\mathrm{i}\vec q \cdot \vec r)P_\infty(r,t) \,\diff^d r$, one can derive a scaling relation for the critical behavior. Hence from \cref{eqscaling_propagator} one concludes
\begin{equation} \label{eqFscaling}
 F_\infty(q,t) = {\cal F}_\infty(q t^{1/d_\text{w}})
\end{equation}
for small wavenumbers $q\ll\sigma^{-1}$ and long times $t\gg t_o$. Substituting $\vec r\mapsto \vec x t^{1/d_\mathrm{w}}$ and $\vec q\mapsto \vec \kappa t^{-1/d_\mathrm{w}}$ in the Fourier transform yields the scaling function as
${\cal F}_\infty( \kappa)=\int \exp\left(\mathrm i \vec \kappa\cdot \vec x\right) {\cal P}_\infty(x)\,\diff^d x$. In particular, the return probability argument determines the large-wavenumber behavior ${\cal  F}_\infty(\kappa\gg 1)\sim \kappa^{-d_\mathrm{f}}$

In the vicinity of the critical density scaling still holds, however, the divergent correlation length $\xi\sim(n^*_c-n^*)^{-\nu}$ has to be introduced~\cite{Kertesz:1983}. The approach towards the scaling law is governed by universal corrections to scaling, which have been worked out recently~\cite{Percolation_EPL:2008}.

\subsection{Mode-coupling theory}
A theory based on first principles has been developed in terms of a self-consistent mode-coupling kinetic theory~\cite{Goetze:1981,Goetze:1981c} for the entire range of densities. Here we focus on the critical regime to rationalize the anticipated scaling behavior. The universal aspects are captured by a simplified approach~\cite{Schnyder:2011} which considers the system composed of a tracer and the porous host structure as a binary mixture, where one component is frozen permanently. The equations of motion then readily follow by specializing the mode-coupling theory of the glass transition to the case considered here.

First, the intermediate scattering function fulfills the exact equation of motion
\begin{align}
 \ddot{F}(q,t) + \nu_s \dot{F}(q,t) + \Omega(q)^2 F(q,t) +  \int_0^t M(q,t-t') \dot{F}(q,t') \diff t' =0
\end{align}
as can be derived within the Zwanzig-Mori projection operator formalism~\cite{Goetze:Complex_Dynamics}. For definiteness we specify the discussion in the remainder of the paper to $d=3$. 
The frequency $\Omega(q)^2 = q^2 v^2/3$ corresponds to the ballistic motion of the particle.
The interaction with the disordered matrix leads to memory effects and friction. The collision rate  $\nu_s = n \pi \sigma^2 v$  induces a Markovian damping and accounts
for uncorrelated scattering with the host matrix, while the non-trivial correlations are hidden in the memory kernel $M(q,t)$.
Within the mode-coupling approximation, the memory kernel is split into a regular smoothly changing background and a mode-coupling contribution that entails a feed-back mechanism. For simplicity we
keep only the latter term, bearing in mind that the Markovian damping should be adjusted to account for the smoothly changing background.
The mode-coupling kernel is a local functional in time of the intermediate scattering functions, linear in the tagged-particle intermediate scattering function. The coupling coefficients referred to as vertices are determined by structural quantities only~\cite{Goetze:Complex_Dynamics}. In addition to the mode-coupling approximation we rely on a generalized hydrodynamics approximation, i.e. we replace the memory kernel by its long-wavelength counterpart $M(q,t) \mapsto M(0,t) =: m(t)$. The microscopic expression is given explicitly by
\begin{equation}
 m(t) =  \frac{n}{3} \int\! \frac{\diff^3 k }{(2\pi)^3 }\, \Omega(k)^2  c_s(k)^2 F(k,t)
\label{eqMCT_kernel}
\end{equation}
For independently distributed scatterers the direct correlation function between matrix and tracer is provided by the spatial Fourier transform of the Mayer function~\cite{Hansen:SimpleLiquids}. 
For hard exclusion one obtains $c_s(k) = -4 \pi  \sigma^2 \text{j}_1(k \sigma)/k$, where $\text{j}_1(\cdot)$ denotes a spherical Bessel function. The critical behavior encoded in this theory coincides with the original self-consistent kinetic theory~\cite{Goetze:1981, Goetze:1981c}.
The mathematical
structure of the simplified mode-coupling theory is identical to the one of the mode-coupling theory of mixtures for which various properties  have been proven rigorously~\cite{Franosch:2002}.
For example, for Brownian dynamics there exist unique solutions with positive spectra, i.e. correlation functions, which are completely monotone functions. 

The MCT approach for the Lorentz model is encoded also in  Krakoviack's MCT description~\cite{Krakoviack:2009} for partially pinned fluids in the limit of a dilute fluid component. There the interaction with the disordered matrix is incorporated on a more sophisticated level than in our equations and enters via connected and disconnected (blocked) parts of the direct correlation function. Our equations coincide with his approach within the generalized hydrodynamics approximation~\cite{Krakoviack:2009}, provided the disconnected correlation function is ignored, which is a common approximation in integral equation theories.

The predictions of the theory for the localization transition can be discussed most conveniently in the Fourier--Laplace domain,  convention $\hat{F}(q,z) = \text{i} \int_0^\infty F(q,t) \text{e}^{\text{i} z t} \diff t$. Then Zwanzig's equation of motion yields
\begin{equation}
 \hat{F}(q,z) = \cfrac{-1}{ z - \cfrac{\Omega(q)^2}{z + \text{i} \nu_s+  \hat{m}(z)}}.
\end{equation}
By linearity $\hat{m}(z)$ is obtained by the same formula as \cref{eqMCT_kernel} with the intermediate scattering functions replaced by their Fourier-Laplace transform $\hat{F}(q,z)$.
Combining both equations leads to single  non-linear equation for the unknown kernel $\hat{m}(z)$. Abbreviating $\hat{\mu}(z) = z + \text{i} \nu_s + \hat{m}(z)$, one has to solve for
\begin{equation}
 z \hat{\mu}(z) - z^2 - \text{i} \nu_s z = \frac{n}{3} \int_0^\infty \frac{4\pi k^2 \diff k}{(2\pi)^3}  c_s(k)^2 \left\{ z \hat{\mu}(z) + \frac{ z^2 \hat{\mu}(z)^2}{\Omega(k)^2 - z\hat{\mu}(z)} \right\}
\end{equation}
A bifurcation occurs at a critical density $n_c^\text{MCT} = 9/4\pi \sigma^3$ where $1 = (n_c^\text{MCT}/3) \int 4\pi k^2 \diff k (2\pi)^{-3}  c_s(k)^2 $. Directly at this critical density
$z \hat{\mu}(z)$ becomes small for frequencies $z\to 0$, and upon expanding one obtains
\begin{equation}
 - z^2 - \text{i} \nu_s z = \frac{n_c^\text{MCT} }{3} \int_0^\infty \frac{4\pi k^2 \diff k}{(2\pi)^3}  c_s(k)^2 \frac{ z^2 \hat{\mu}(z)^2}{\Omega(k)^2} + {\cal O}( z \hat{\mu}(z))^{5/2}
\end{equation}
where the correction term arises from the singular behavior of the denominator for long wavelength. The integral can be performed yielding $-\text{i} z \nu_s = (6 \sigma^2 / 5 v^2) [-z \hat{m}(z)]^2$ as leading contribution to the equation. Thus at the  critical point the memory kernel displays singular behavior $-z \hat{m}(z) = (- \text{i}  z   t_s)^{1/2} v^2 /3 \sigma^2   + {\cal O}(z)$ with $t_s = 135 \sigma /8 v$ as has been calculated already by G\"otze \emph{et al}~\cite{Goetze:1981,Goetze:1981c}. This implies for the intermediate scattering function for long-wavelengths and low frequencies
\begin{equation}
-z \hat{F}(q,z) \to   \frac{1}{1+ q^2 \sigma^2 (-\text{i} z t_s)^{-1/2} }
\end{equation}
The right-hand side displays spatio-temporal scaling behavior, it depends only on the combination $z/q^{4}$ which implies a walk dimension $d_\text{w}^\text{MCT}=4$. Furthermore for large  wavenumbers
$q \sigma \gg ( |z| t_s )^{-1/4}$, the scaling function decays as $q^{-2}$ from which we infer the fractal dimension $d_\text{f}^\text{MCT} =2$. Transforming back to the temporal domain,
we infer that at the critical density, $F(q,t)$ displays scaling behavior for long times and small wavenumbers similar to \cref{eqFscaling} with exponents that are simple numbers.

The precise value of the critical density within mode-coupling theory is determined by the properties of the vertex, i.e. the coupling coeffients. In the standard factorization procedure used here, $n_c^\text{MCT} \sigma^3=9/4\pi \approx 0.716$ is surprisingly close to the percolation threshold $n_c \sigma^3 \approx 0.838$. 
The situation is less favorable in two dimensions, where MCT yields a value of 
$2/\pi \approx 0.637$ whereas the percolation occurs at 0.359. Interestingly, if one extrapolates the rigorous low-density expansion for the case of Brownian tracers~\cite{Lorentz_VACF:2010}, the situation is reversed. There the  low-density expansion leads to $3/2 \pi \approx0.48$ in three dimension  whereas in two dimensions one arrives at $1/\pi \approx 0.318$. 

Strictly speaking the scaling properties discussed above are a consequence of the generalized hydrodynamics approximation. Using the full wavevector-dependent theory~\cite{Schnyder:2011} these exponents hold only in an intermediate regime. This phenomenon of spurious long-wavelength singularities in MCT approaches is now well-understood~\cite{Schnyder:2011}  and we do not repeat the discussion here. Ultimately it remains unclear how to properly handle the emergence of fractal structures and the divergent length scale~\cite{Goetze:Complex_Dynamics}, nevertheless we shall show that certain aspects predicted by the theory have a close counterpart in the simulation results.

Rather than relying on asymptotic analysis we solved the mode-coupling equations numerically in the temporal domain. The integral in the memory kernel is one-dimensional in the wavenumber $k$ when expressed in spherical coordinates. To account for the long-wavelength behavior we have used a wave\-number discretization with $N_\mathrm{s}$ logarithmically-spaced wave\-numbers from a smallest $k_{\mathrm{min}}$ up to a wavenumber $\Delta k$. From there we use an equidistant grid with spacing $\Delta k$ up to a largest wavenumber $k_{\mathrm{max}}$. We used $N_\mathrm{s}=50$,  $k_{\mathrm{min}}=10^{-8}/R$, $\Delta k = 0.4/R$ and $k_{\mathrm{max}}=24/R$. With these parameters, the critical density is slightly shifted to $n^*_c\approx 0.748$ but converges to the exact value on finer discretizations. The singular behavior of the memory kernel given above matches the numerical results to within $1\%$ of its amplitude at long times.

A standard MCT algorithm was used to solve the equation on an equidistant time grid, which is repeatedly coarsened by factors of 2 in order to be able to calculate the solution over many decades in time. The memory kernel was numerically integrated on the wavenumber grid with the trapezoidal rule.

We have checked that the spurious long-wavelength singularities do not introduce a  qualitative difference  in the ranges displayed in the figures, with the exception of the mean-square displacement (not shown), by comparing to the full wavenumber-dependent theory. For a detailed discussion of the implications of using a discretized  wavenumber-dependent mode-coupling functional we refer 
to  Schnyder et al~\cite{Schnyder:2011}. 

\section{Simulation}

We generate trajectories by event-driven molecular dynamics simulations at the critical density  for system sizes $L=200\sigma$ corresponding to 6,704,000 obstacles.
We impose periodic boundary conditions to minimize finite-size effects.
 Initial particle positions are sampled only on the remaining infinite cluster, which we identify by a numerical Voronoi tessellation~\footnote{We have employed the free \texttt{voro++} package originally developed by Chris
Rycroft, see \texttt{http://math.lbl.gov/voro++/}.}. Out of 96 realizations of the disorder only 44 displayed an infinite cluster, consistent with the theoretical expectation that in finite systems the transition is smeared. We extract a critical density~\cite{Spanner:thesis} $n_c^*=0.838$ which is slightly lower than the value $0.839$ found for $L=200\sigma$ in Refs. \cite{Lorentz_PRL:2006, Lorentz_JCP:2008}. The difference can be rationalized by the bias to obtain half of the samples to be percolating.

Correlation functions are obtained by moving-time averages for lag times up to $3\cdot10^9t_o$ using a standard blocking scheme~\cite{Colberg:2011} where the trajectory is sampled essentially on a logarithmic time grid. Our results are extracted from 7 independent trajectories in each of the 44 percolating configurations. We measure the incoherent intermediate scattering function directly as defined in \cref{eqdef_isf} with the constraint that the wave vectors are consistent with the periodic boundaries. The smallest wavenumber that can be achieved in the simulation is limited by the system size to $2\pi/L$, i.e. $q=0.031\sigma^{-1}$.

The main computational effort lies in the identification of the percolating cluster and to locate the critical
density with sufficient accuracy.
The limitation of the system size is set by computer memory necessary (25\,GB) to analyze the Voronoi network. Once the infinite cluster is identified, a single trajectory requires typically 12 hours on a prevalent 3\,GHz CPU, resulting in 150 days of computing time per system size and density.

\section{Results and discussion}

\begin{figure}
\includegraphics[angle=0,width=0.9\linewidth]{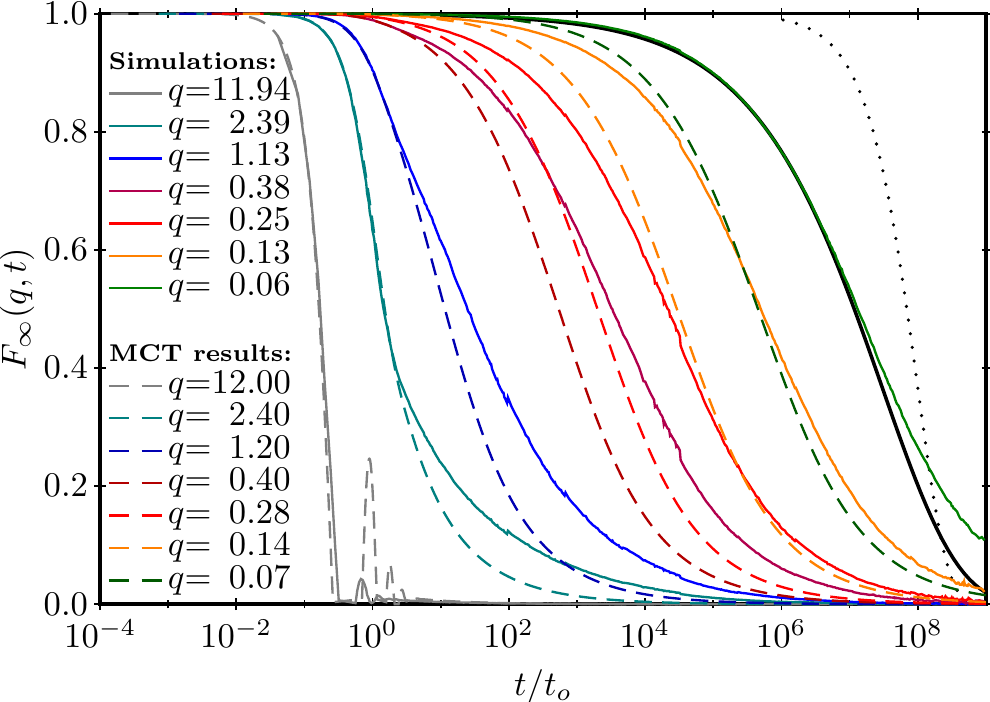}
\caption{Intermediate scattering function $F_\infty(q,t)$ for tracers exploring only the percolating cluster. Time is measured in units of $t_o=\sigma/v$. The simulation is performed  directly at the critical density $n_c^*=0.838$ and compared to the mode coupling theory. The dotted black  line corresponds to an exponential relaxation for comparison. The full black line is a fit to a stretched exponential $\exp(-(t/\tau_q)^\beta)$, $\beta=0.39$.
}
\label{figisf_raw}
\end{figure}

Simulation results for the three-dimensional Lorentz model have been presented mostly in terms of the mean-square displacement\cite{Lorentz_PRL:2006,Lorentz_JCP:2008,Spanner:2011}. Spatio-temporal information beyond the second moment is encoded in the intermediate scattering function, as shown in \cref{figisf_raw} for the obstacle density directly at the transition point. The time window in the simulation covers 9 non-trivial decades and the data displays practically no noise. The range of wavenumbers includes a factor of almost 200 corresponding to the smallest length scale of an obstacle diameter $\sigma$ to the box size $L=200\sigma$. A pronounced bump  becomes apparent at the collision time $t_c\approx t_o$ for wavenumbers probing the microscopic scales, familiar from the dynamics of dense liquids~\cite{Hansen:SimpleLiquids,Goetze:Complex_Dynamics}. For long wavelengths the decay becomes considerably non-exponential, rather the shape is characterized by significant stretching. A fit to a  Kohlrausch-Williams-Watts function $\exp(-(t/\tau_q)^\beta)$ provides a nice description of the data down to a value of 0.4 and yields a rather low stretching exponent $\beta=0.39$. For even lower values of $q$, the stretching becomes even more significant. Within the long wavelength regime the shape of the ISF become insensitive to changes in $q$, only the characteristic time scale varies.
The MCT results for the critical obstacle density $n_c^\text{MCT}$ are included in the figure
for comparison. For moderately small wavenumbers, $0.4\lesssim q\lesssim 2.5$,  the agreement is reasonable, although a pronounced tail appears in the simulation data that is not captured by the theory. For lower wavenumbers the shape of the relaxation curves is still similar to the simulation data, yet the time scales start to deviate significantly.

\begin{figure}
\includegraphics[angle=0,width=0.9\linewidth]{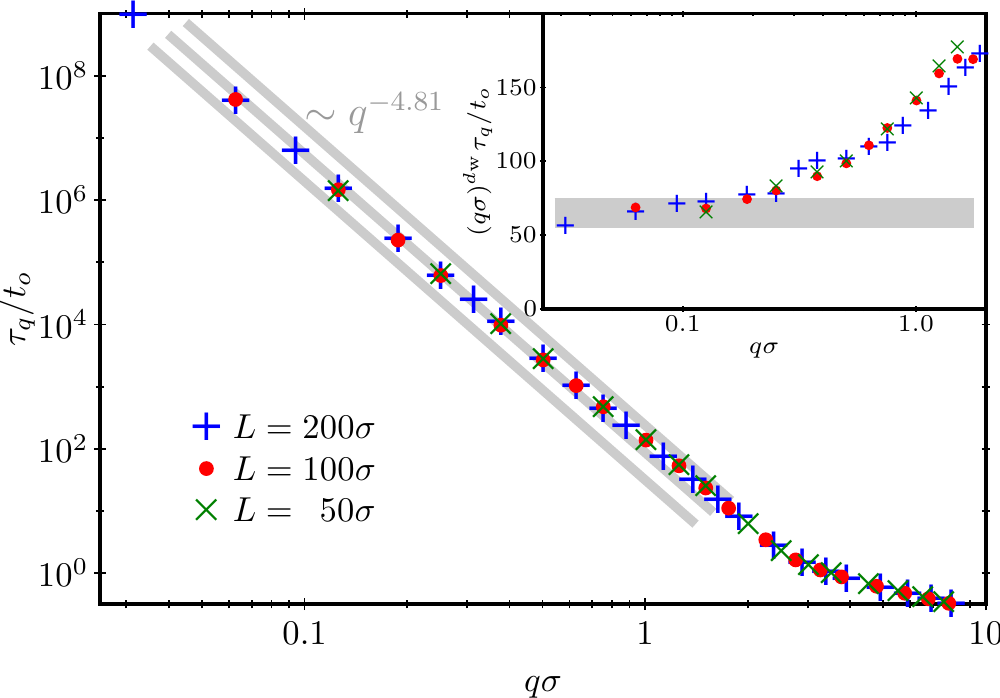}
\caption{Relaxation time $\tau_q$ of the simulated intermediate scattering function at the critical density $n_c^*=0.838$ for different wave vectors $q$
and system sizes $L$. The thick grey lines indicate a power law expected from the scaling hypothesis and serves as a guide to the eye. Inset: Rectification plot of the relaxation times testing the asymptotic power-law behavior. }
\label{figisf_tauq}
\end{figure}

Decay times $\tau_q$ defined by $F_\infty(q,\tau_q)=1/\mathrm{e}$ vary over 8 orders of magnitude. Note that for ordinary diffusion in the hydrodynamic regime, one would expect a change of $\tau_q$ by only 4 orders of magnitude, since then $F(q,t)=\exp(-Dq^2t)$ and $\tau_q\propto q^{-2}$.
The relaxation times $\tau_q$ as a function of $q$ are exhibited in \cref{figisf_tauq} using a double-logarithmic representation.
The data approach a power law $\tau_q\sim q^{-d_\mathrm{w}}$ in the regime of small wavenumbers and, in principle, 
 allow the determination of the walk dimension $d_\mathrm{w}$.
The rectification plot (inset of \cref{figisf_tauq}) illustrates that the power-law behavior is approached only asymptotically. We fix the scale factor in $\tau_q/t_o \simeq A_\tau (q\sigma)^{-d_\text{w}}$ to $A_\tau = 65\pm10$.
From the slope one can in principle obtain the value for $d_\text{w}$ which 
 is compatible with our earlier results~\cite{Spanner:2011,Franosch:2011} extracted from mean-square displacement $\delta r^2_\infty(t)\sim t^{2/d_\mathrm{w}}$.
Of course, $d_\mathrm{w}$ can be determined more accurately from $\delta r^2_\infty(t)$ since the time window spans 8 decades, whereas the $q$-range is varied only by a factor of 200, hence we rely on our earlier estimate in the remainder of this article.

\begin{figure}
\includegraphics[angle=0,width=0.9\linewidth]{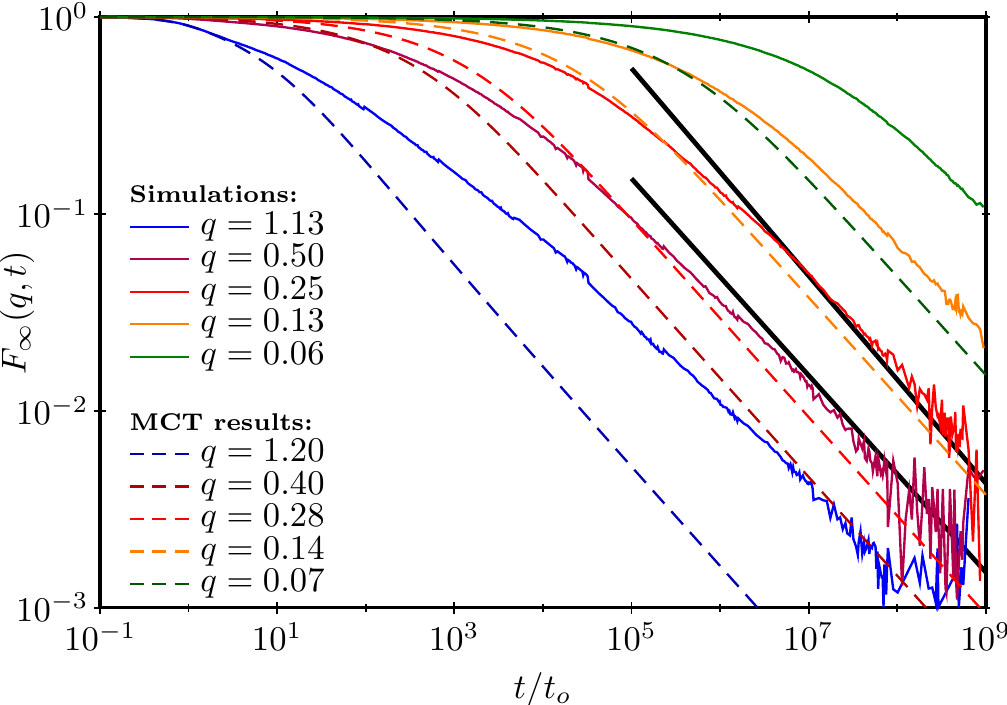}
\caption{Intermediate scattering function at the critical density, double-logarithmic representation. The terminal relaxation approaches a power law for small wavenumbers.
As guide to the eye: solid black lines are power laws $t^{-d_\mathrm{f}/d_\mathrm{w}}$ and $t^{-1/2}$ respectively.} \label{figisf_log}
\end{figure}

To investigate the behavior at long times we display the simulation data in a double-logarithmic representation in \cref{figisf_log}. In the long-wavelength regime the curves approach a power-law relaxation for $t\to\infty$. From the scaling theory at the critical point, \cref{eqFscaling}, for small rescaled arguments one anticipates
$F_\infty(q,t)\sim t^{-d_\mathrm{s}/2},  t\to\infty$, where $d_\mathrm{s}=2d_\mathrm{f}/d_\mathrm{w}$ is known as the spectral dimension~\cite{benAvraham:DiffusionInFractals}. Our data are compatible with the value $d_\mathrm{s}=1.054$ obtained from the previously determined numbers for $d_\mathrm{f}$ and $d_\mathrm{w}$. The spectral dimension for three-dimensional Lorentz models is surprisingly close to the MCT result $d^\mathrm{MCT}_\mathrm{s}=1$, in fact based on the window accessible in today's computer simulations these values are indistinguishable.

\begin{figure}
\includegraphics[angle=0,width=0.9\linewidth]{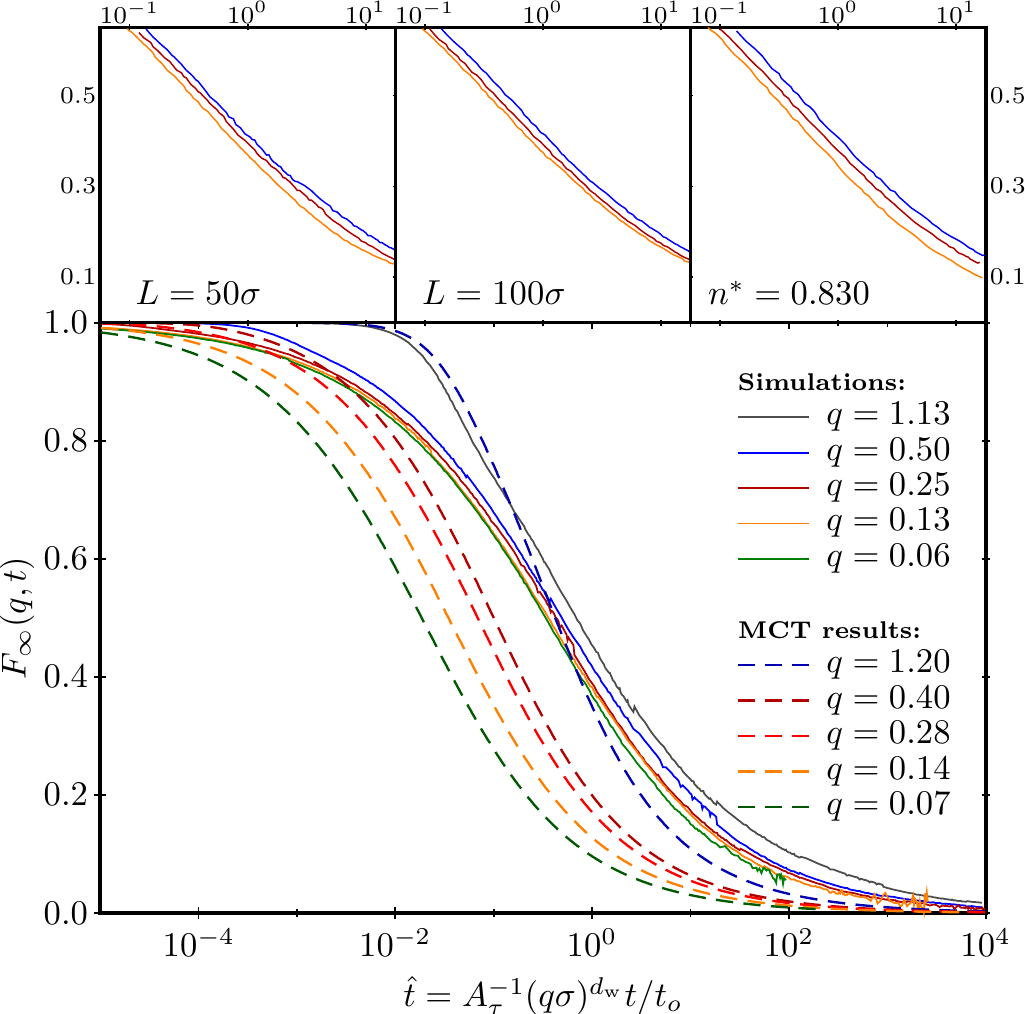}
\caption{ISF rescaled as a function of rescaled time $\hat t = A_\tau^{-1} (q\sigma)^{d_\mathrm{w}}t/t_o$ for small wavenumbers $q$. The lower panel corresponds to the largest system size investigated ($L=200\sigma$)  at  the critical density ($n*=0.838$). The top panels are for smaller system sizes (left, middle), or larger distance ($(n_c*-n^*)/n_c^*=0.01$) to the critical density\label{figisf_scale}. 
}
\end{figure}

In addition to reading off the critical exponents from relaxation times $\tau_q$ and the power-law terminal relaxation, we investigate the scaling properties, see \cref{figisf_scale}. Shifting the data according to rescaled time $\hat t \sim q^{d_\mathrm{w}}t$ the curves almost superimpose. Upon decreasing the wave number, the data collapse improves and asymptotically a universal master curve is approached. Deviations at moderately small wave numbers are predominantly due to the poor convergence of $\tau_q/q^{-d_\mathrm{w}}$ (\cref{figisf_tauq}) whereas the shape appears to be approached more rapidly. In fact using the measured $\tau_q$ rather than its asymptotic prediction enforces data collapse at the point $(1, \mathrm{e}^{-1})$  and the scaling function is nicely followed.
The scaling plot is sensitive to finite size corrections and fine-tuning to the critical density, as shown in the top panels of \cref{figisf_scale}.
Interestingly, MCT displays an approximate data collapse for small wavenumbers, as is explained by the small difference in $d_\mathrm{w}$.

\begin{figure}[htp]
\includegraphics[angle=0,width=0.9\linewidth]{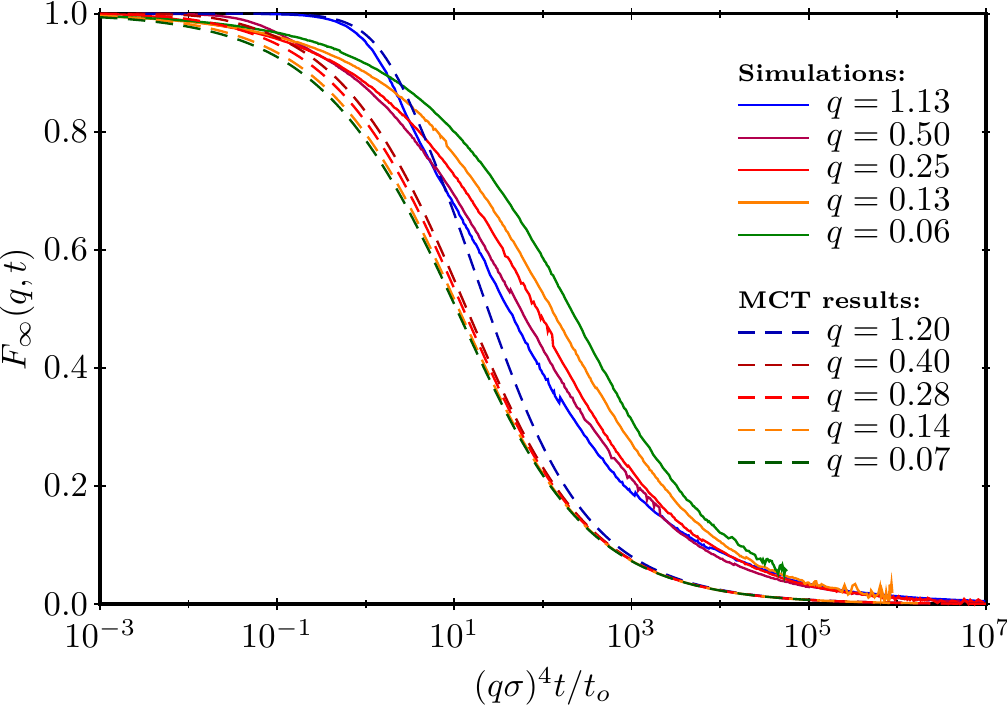}
\caption{For comparison: rescaling of the intermediate scattering function with rescaled time $ (q \sigma)^4 t/t_o$ as suggested by MCT for the same data as in Fig.~\ref{figisf_raw}.
} \label{figisf_scale_mct}
\end{figure}

The MCT estimate for the walk dimension is $d_\mathrm{w}^\mathrm{MCT}=4$ is sufficiently close to the measured value $d_\mathrm{w}=4.81$ to analyze the data with a rescaled time $\hat{t} 
\sim q^{4}t$. Such a rescaling is displayed in \cref{figisf_scale_mct}, and indeed the simulation data show reasonable data collapse, given that the original data exhibit a spread of six decades. As expected, the numerical solutions of the MCT equations now collapse perfectly for small $q$.

\begin{figure}[htp]
\includegraphics[angle=0,width=0.9\linewidth]{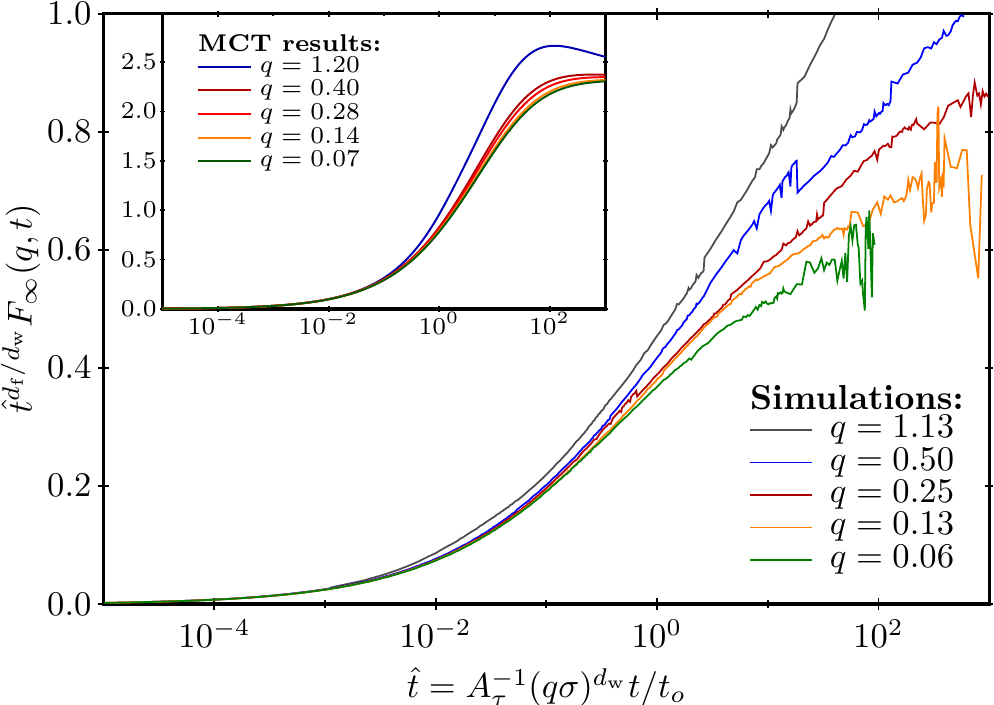}
\caption{Rescaled and rectification highlights the approach to a master curve. The scaling variable for the simulation data is $\hat{t} = A_\tau^{-1}  (t/t_o) (q \sigma)^{d_\text{w}}$ and spectral dimension $d_s = 1.054$,  
respectively (in the inset) $\hat{t} = (t/t_o) (q\sigma)^4$ and spectral dimension $1$.
\label{figisf_rect}}
\end{figure}

Since the scaling prediction is an asymptotic law, valid for long times and small wave numbers, an even more sensitive test is obtained by plotting $\hat t^{d_\mathrm{f}/d_\mathrm{w}}F_\infty(q,t)$ vs. rescaled time $\hat t \sim q^{d_\mathrm{w}}t$. Then the long-time behavior of $\hat F_\infty$ is highlighted and the approach towards a power-law relaxation is mapped to a finite limit. Our simulation data (see \cref{figisf_rect}) nicely follow the expected scaling for small $\hat t$; for long times, however, deviations become visible. Nevertheless, one infers slow convergence for small wavenumbers $q\to0$. In principle, the scaling hypothesis can be extended to include universal corrections to scaling~\cite{Percolation_EPL:2008} to rationalize the residual spread.

The success of the theory is encouraging on the one hand, but delusive upon closer inspection. First, we chose to compare only the motion on the infinite cluster to MCT, rather than the all-cluster average dynamics. While a dynamic scaling hypothesis can be worked out based on the percolation picture, see e.g.\ Refs.~\citenum{Kertesz:1984} and \citenum{Percolation_EPL:2008}, the exponents start to deviate significantly from the measured ones. Second, there is a conceptual problem, since MCT does not distinguish between the two types of tracers. While the tracers on the infinite cluster are ergodic, compatible with the MCT notion,
the all-cluster average motion is not, since some particles will always be trapped in finite pockets of the host structure. These holes are present at arbitrarily low obstacle density, leading to non-vanishing long-time limits of the ISF sensitively dependent of wavenumber~\cite{Franosch:2011}. This peculiarity supports the idea that we have pursued in the current paper to make the comparison with the infinite cluster motion only. Yet, for obstacle densities above the transition,  MCT predicts a continuous increase of the long-time limits of the ISF. Since the infinite cluster ceases to exist a comparison here is excluded.

\section{Summary and Conclusion}
We have measured the intermediate scattering function for a three-dimensional Lorentz model directly at the percolation transition for particles confined to the infinite cluster. The dynamics becomes anomalous  due to the underlying spatial fractal and a scaling hypothesis is expected to hold. We have tested various aspects of spatio-temporal transport thereby corroborating the notion of self-similarity  beyond the second moment. In particular, we find a power-law relaxation in the long-wavelength regime strikingly different from the two-step relaxation scenario known from glasses and supercooled liquids~\cite{Goetze:Complex_Dynamics}. We have compared our simulation results within  a simplified mode-coupling theory that reproduces the critical properties of a self-consistent mode-coupling kinetic theory~\cite{Goetze:1981,Goetze:1981c}. The exponents predicted by this approach are sufficiently close to the measured values to allow for a quantitative test of the theory and surprisingly good agreement is found.

While the agreement with MCT should not be overstated for the reasons
discussed above, it does highlight that the theory continues to provide
qualitative, and sometimes semi-quantitative, predictions also in
the vicinity of certain percolation-type transitions. This is of importance,
since there is now emerging a renewed interest in anomalous transport and its
interplay with glass-transition phenomena, for example in the case of systems
with self-generated disorder (such as size-disparate binary mixtures).
Since true asymptotic scaling behavior emerges only rather close to the
transition and often outside experimentally accessible windows, MCT
proves useful as a theory that can also, beyond scaling analysis, predict
generic (non-universal, but typical) behavior. 

While we have clarified what the intermediate scattering functions should look like, it remains a challenge for the future to derive a mode-coupling theory from first principles that accounts for the percolation transition. A possible route would be to build the theory on the obstacle clusters that form and use renormalized vertices where the geometry of the obstacle parcours is reflected properly. The idea for such a cluster-MCT has been promoted for the problem of  weak gelation of attractive colloids~\cite{Kroy:2004}, where colloids stick to clusters moving together as entities.
However, it remains open how such a theory  can be achieved in practice. A second promising approach is to employ a diagrammatic expansion~\cite{Szamel:2007} for the dynamics to first rederive the simplified MCT equations and then identify what processes have to be included to account for the underlying fractal geometry.

\section{Acknowledgement}
We thank V. Krakoviack for insightful remarks.
This project has been supported via the DFG research unit FOR-1394 `Nonlinear response to probe vitrification', project P8. Th.~V.\ thanks for funding from the Helmholtz-Gemeinschaft (HGF VH-NG 406), and the Zukunftskolleg of the University of Konstanz.

\providecommand*{\mcitethebibliography}{\thebibliography}
\csname @ifundefined\endcsname{endmcitethebibliography}
{\let\endmcitethebibliography\endthebibliography}{}


\begin{mcitethebibliography}{53}
\providecommand*{\natexlab}[1]{#1}
\providecommand*{\mciteSetBstSublistMode}[1]{}
\providecommand*{\mciteSetBstMaxWidthForm}[2]{}
\providecommand*{\mciteBstWouldAddEndPuncttrue}
  {\def\EndOfBibitem{\unskip.}}
\providecommand*{\mciteBstWouldAddEndPunctfalse}
  {\let\EndOfBibitem\relax}
\providecommand*{\mciteSetBstMidEndSepPunct}[3]{}
\providecommand*{\mciteSetBstSublistLabelBeginEnd}[3]{}
\providecommand*{\EndOfBibitem}{}
\mciteSetBstSublistMode{f}
\mciteSetBstMaxWidthForm{subitem}
{(\emph{\alph{mcitesubitemcount}})}
\mciteSetBstSublistLabelBeginEnd{\mcitemaxwidthsubitemform\space}
{\relax}{\relax}

\bibitem[Angell(1995)]{Angell:1995}
C.~Angell, \emph{Science}, 1995, \textbf{267}, 1924--1935\relax
\mciteBstWouldAddEndPuncttrue
\mciteSetBstMidEndSepPunct{\mcitedefaultmidpunct}
{\mcitedefaultendpunct}{\mcitedefaultseppunct}\relax
\EndOfBibitem
\bibitem[G\"otze and Sj\"ogren({1992})]{Goetze:1992}
W.~G\"otze and L.~Sj\"ogren, \emph{{Rep. Prog. Phys.}}, {1992}, \textbf{{55}},
  {241--376}\relax
\mciteBstWouldAddEndPuncttrue
\mciteSetBstMidEndSepPunct{\mcitedefaultmidpunct}
{\mcitedefaultendpunct}{\mcitedefaultseppunct}\relax
\EndOfBibitem
\bibitem[G\"otze({1999})]{Goetze:1999}
W.~G\"otze, \emph{{J. Phys. Condens. Matter}}, {1999}, \textbf{{11}},
  {A1--A45}\relax
\mciteBstWouldAddEndPuncttrue
\mciteSetBstMidEndSepPunct{\mcitedefaultmidpunct}
{\mcitedefaultendpunct}{\mcitedefaultseppunct}\relax
\EndOfBibitem
\bibitem[Pusey and van Megen(1986)]{Pusey:1986}
P.~N. Pusey and W.~van Megen, \emph{Nature}, 1986, \textbf{320}, 340
  --342\relax
\mciteBstWouldAddEndPuncttrue
\mciteSetBstMidEndSepPunct{\mcitedefaultmidpunct}
{\mcitedefaultendpunct}{\mcitedefaultseppunct}\relax
\EndOfBibitem
\bibitem[G\"otze(2009)]{Goetze:Complex_Dynamics}
W.~G\"otze, \emph{Complex Dynamics of Glass-Forming Liquids -- A Mode-Coupling
  Theory}, Oxford, Oxford, 2009\relax
\mciteBstWouldAddEndPuncttrue
\mciteSetBstMidEndSepPunct{\mcitedefaultmidpunct}
{\mcitedefaultendpunct}{\mcitedefaultseppunct}\relax
\EndOfBibitem
\bibitem[Krakoviack(2005)]{Krakoviack:2005}
V.~Krakoviack, \emph{Phys. Rev. Lett.}, 2005, \textbf{94}, 065703\relax
\mciteBstWouldAddEndPuncttrue
\mciteSetBstMidEndSepPunct{\mcitedefaultmidpunct}
{\mcitedefaultendpunct}{\mcitedefaultseppunct}\relax
\EndOfBibitem
\bibitem[Krakoviack(2007)]{Krakoviack:2007}
V.~Krakoviack, \emph{Phys. Rev. E}, 2007, \textbf{75}, 031503\relax
\mciteBstWouldAddEndPuncttrue
\mciteSetBstMidEndSepPunct{\mcitedefaultmidpunct}
{\mcitedefaultendpunct}{\mcitedefaultseppunct}\relax
\EndOfBibitem
\bibitem[Krakoviack(2009)]{Krakoviack:2009}
V.~Krakoviack, \emph{Phys. Rev. E}, 2009, \textbf{79}, 061501\relax
\mciteBstWouldAddEndPuncttrue
\mciteSetBstMidEndSepPunct{\mcitedefaultmidpunct}
{\mcitedefaultendpunct}{\mcitedefaultseppunct}\relax
\EndOfBibitem
\bibitem[Krakoviack(2011)]{Krakoviack:2011}
V.~Krakoviack, \emph{Phys. Rev. E}, 2011, \textbf{84}, 050501\relax
\mciteBstWouldAddEndPuncttrue
\mciteSetBstMidEndSepPunct{\mcitedefaultmidpunct}
{\mcitedefaultendpunct}{\mcitedefaultseppunct}\relax
\EndOfBibitem
\bibitem[Kurzidim \emph{et~al.}(2009)Kurzidim, Coslovich, and
  Kahl]{Kurzidim:2009}
J.~Kurzidim, D.~Coslovich and G.~Kahl, \emph{Phys. Rev. Lett.}, 2009,
  \textbf{103}, 138303\relax
\mciteBstWouldAddEndPuncttrue
\mciteSetBstMidEndSepPunct{\mcitedefaultmidpunct}
{\mcitedefaultendpunct}{\mcitedefaultseppunct}\relax
\EndOfBibitem
\bibitem[Kurzidim \emph{et~al.}({2010})Kurzidim, Coslovich, and
  Kahl]{Kurzidim:2010}
J.~Kurzidim, D.~Coslovich and G.~Kahl, \emph{{Phys. Rev. E}}, {2010},
  \textbf{{82}}, 041505\relax
\mciteBstWouldAddEndPuncttrue
\mciteSetBstMidEndSepPunct{\mcitedefaultmidpunct}
{\mcitedefaultendpunct}{\mcitedefaultseppunct}\relax
\EndOfBibitem
\bibitem[Kurzidim \emph{et~al.}({2011})Kurzidim, Coslovich, and
  Kahl]{Kurzidim:2011}
J.~Kurzidim, D.~Coslovich and G.~Kahl, \emph{{J. Phys.: Condens. Matter}},
  {2011}, \textbf{{23}}, 234122\relax
\mciteBstWouldAddEndPuncttrue
\mciteSetBstMidEndSepPunct{\mcitedefaultmidpunct}
{\mcitedefaultendpunct}{\mcitedefaultseppunct}\relax
\EndOfBibitem
\bibitem[Kim \emph{et~al.}(2009)Kim, Miyazaki, and Saito]{Kim:2009}
K.~Kim, K.~Miyazaki and S.~Saito, \emph{Europhys. Lett. (EPL)}, 2009,
  \textbf{88}, 36002\relax
\mciteBstWouldAddEndPuncttrue
\mciteSetBstMidEndSepPunct{\mcitedefaultmidpunct}
{\mcitedefaultendpunct}{\mcitedefaultseppunct}\relax
\EndOfBibitem
\bibitem[Kim \emph{et~al.}({2011})Kim, Miyazaki, and Saito]{Kim:2010}
K.~Kim, K.~Miyazaki and S.~Saito, \emph{{J. Phys.: Condens. Matter}}, {2011},
  \textbf{{23}}, {234123}\relax
\mciteBstWouldAddEndPuncttrue
\mciteSetBstMidEndSepPunct{\mcitedefaultmidpunct}
{\mcitedefaultendpunct}{\mcitedefaultseppunct}\relax
\EndOfBibitem
\bibitem[Kim \emph{et~al.}({2010})Kim, Miyazaki, and Saito]{Kim:2010a}
K.~Kim, K.~Miyazaki and S.~Saito, \emph{{Eur. Phys. J. Special Topics}},
  {2010}, \textbf{{189}}, {135--139}\relax
\mciteBstWouldAddEndPuncttrue
\mciteSetBstMidEndSepPunct{\mcitedefaultmidpunct}
{\mcitedefaultendpunct}{\mcitedefaultseppunct}\relax
\EndOfBibitem
\bibitem[Gallo \emph{et~al.}({2009})Gallo, Attili, and Rovere]{Gallo:2009}
P.~Gallo, A.~Attili and M.~Rovere, \emph{{Phys. Rev. E}}, {2009},
  \textbf{{80}}, 061502\relax
\mciteBstWouldAddEndPuncttrue
\mciteSetBstMidEndSepPunct{\mcitedefaultmidpunct}
{\mcitedefaultendpunct}{\mcitedefaultseppunct}\relax
\EndOfBibitem
\bibitem[Gallo and Rovere({2011})]{Gallo:2011}
P.~Gallo and M.~Rovere, \emph{{J. Phys. Condens. Matter}}, {2011},
  \textbf{{23}}, {234118}\relax
\mciteBstWouldAddEndPuncttrue
\mciteSetBstMidEndSepPunct{\mcitedefaultmidpunct}
{\mcitedefaultendpunct}{\mcitedefaultseppunct}\relax
\EndOfBibitem
\bibitem[Horbach \emph{et~al.}(2002)Horbach, Kob, and Binder]{Horbach:2002}
J.~Horbach, W.~Kob and K.~Binder, \emph{Phys. Rev. Lett.}, 2002, \textbf{88},
  125502\relax
\mciteBstWouldAddEndPuncttrue
\mciteSetBstMidEndSepPunct{\mcitedefaultmidpunct}
{\mcitedefaultendpunct}{\mcitedefaultseppunct}\relax
\EndOfBibitem
\bibitem[Meyer \emph{et~al.}(2004)Meyer, Horbach, Kob, Kargl, and
  Schober]{Meyer:2004}
A.~Meyer, J.~Horbach, W.~Kob, F.~Kargl and H.~Schober, \emph{Phys. Rev. Lett.},
  2004, \textbf{93}, 027801\relax
\mciteBstWouldAddEndPuncttrue
\mciteSetBstMidEndSepPunct{\mcitedefaultmidpunct}
{\mcitedefaultendpunct}{\mcitedefaultseppunct}\relax
\EndOfBibitem
\bibitem[Voigtmann and Horbach(2006)]{Voigtmann:2006}
{\relax Th}.~Voigtmann and J.~Horbach, \emph{Europhys. Lett.}, 2006,
  \textbf{74}, 459--465\relax
\mciteBstWouldAddEndPuncttrue
\mciteSetBstMidEndSepPunct{\mcitedefaultmidpunct}
{\mcitedefaultendpunct}{\mcitedefaultseppunct}\relax
\EndOfBibitem
\bibitem[Moreno and Colmenero(2006)]{Moreno:2006}
A.~J. Moreno and J.~Colmenero, \emph{J. Chem. Phys.}, 2006, \textbf{125},
  164507\relax
\mciteBstWouldAddEndPuncttrue
\mciteSetBstMidEndSepPunct{\mcitedefaultmidpunct}
{\mcitedefaultendpunct}{\mcitedefaultseppunct}\relax
\EndOfBibitem
\bibitem[Moreno and Colmenero(2006)]{Moreno:2006a}
A.~J. Moreno and J.~Colmenero, \emph{Phys. Rev. E}, 2006, \textbf{74},
  021409\relax
\mciteBstWouldAddEndPuncttrue
\mciteSetBstMidEndSepPunct{\mcitedefaultmidpunct}
{\mcitedefaultendpunct}{\mcitedefaultseppunct}\relax
\EndOfBibitem
\bibitem[Voigtmann and Horbach(2009)]{Voigtmann:2009}
{\relax Th}.~Voigtmann and J.~Horbach, \emph{Phys. Rev. Lett.}, 2009,
  \textbf{103}, 205901\relax
\mciteBstWouldAddEndPuncttrue
\mciteSetBstMidEndSepPunct{\mcitedefaultmidpunct}
{\mcitedefaultendpunct}{\mcitedefaultseppunct}\relax
\EndOfBibitem
\bibitem[Voigtmann(2011)]{Voigtmann:2011}
{\relax Th}.~Voigtmann, \emph{EPL}, 2011, \textbf{96}, 36006\relax
\mciteBstWouldAddEndPuncttrue
\mciteSetBstMidEndSepPunct{\mcitedefaultmidpunct}
{\mcitedefaultendpunct}{\mcitedefaultseppunct}\relax
\EndOfBibitem
\bibitem[Kikuchi and Horbach(2007)]{Kikuchi:2007}
N.~Kikuchi and J.~Horbach, \emph{Europhys. Lett. (EPL)}, 2007, \textbf{77},
  2601\relax
\mciteBstWouldAddEndPuncttrue
\mciteSetBstMidEndSepPunct{\mcitedefaultmidpunct}
{\mcitedefaultendpunct}{\mcitedefaultseppunct}\relax
\EndOfBibitem
\bibitem[H{\"o}f\/ling \emph{et~al.}(2006)H{\"o}f\/ling, Franosch, and
  Frey]{Lorentz_PRL:2006}
F.~H{\"o}f\/ling, T.~Franosch and E.~Frey, \emph{Phys. Rev. Lett.}, 2006,
  \textbf{96}, 165901\relax
\mciteBstWouldAddEndPuncttrue
\mciteSetBstMidEndSepPunct{\mcitedefaultmidpunct}
{\mcitedefaultendpunct}{\mcitedefaultseppunct}\relax
\EndOfBibitem
\bibitem[H{\"o}f\/ling \emph{et~al.}(2008)H{\"o}f\/ling, Munk, Frey, and
  Franosch]{Lorentz_JCP:2008}
F.~H{\"o}f\/ling, T.~Munk, E.~Frey and T.~Franosch, \emph{J. Chem. Phys.},
  2008, \textbf{128}, 164517\relax
\mciteBstWouldAddEndPuncttrue
\mciteSetBstMidEndSepPunct{\mcitedefaultmidpunct}
{\mcitedefaultendpunct}{\mcitedefaultseppunct}\relax
\EndOfBibitem
\bibitem[Bauer \emph{et~al.}({2010})Bauer, H\"ofling, Munk, Frey, and
  Franosch]{Bauer:2010}
T.~Bauer, F.~H\"ofling, T.~Munk, E.~Frey and T.~Franosch, \emph{{Eur. Phys. J.
  Special Topics}}, {2010}, \textbf{{189}}, {103--118}\relax
\mciteBstWouldAddEndPuncttrue
\mciteSetBstMidEndSepPunct{\mcitedefaultmidpunct}
{\mcitedefaultendpunct}{\mcitedefaultseppunct}\relax
\EndOfBibitem
\bibitem[Franosch \emph{et~al.}({2011})Franosch, Spanner, Bauer,
  Schr{\"o}der-Turk, and H{\"o}fling]{Franosch:2011}
T.~Franosch, M.~Spanner, T.~Bauer, G.~E. Schr{\"o}der-Turk and F.~H{\"o}fling,
  \emph{{J. Non.-Cryst. Solids}}, {2011}, \textbf{{357}}, {472--478}\relax
\mciteBstWouldAddEndPuncttrue
\mciteSetBstMidEndSepPunct{\mcitedefaultmidpunct}
{\mcitedefaultendpunct}{\mcitedefaultseppunct}\relax
\EndOfBibitem
\bibitem[Spanner \emph{et~al.}({2011})Spanner, H\"ofling, Schr\"oder-Turk,
  Mecke, and Franosch]{Spanner:2011}
M.~Spanner, F.~H\"ofling, G.~E. Schr\"oder-Turk, K.~Mecke and T.~Franosch,
  \emph{{J. Phys.: Condens. Matter}}, {2011}, \textbf{{23}}, {234120}\relax
\mciteBstWouldAddEndPuncttrue
\mciteSetBstMidEndSepPunct{\mcitedefaultmidpunct}
{\mcitedefaultendpunct}{\mcitedefaultseppunct}\relax
\EndOfBibitem
\bibitem[G\"otze \emph{et~al.}({1981})G\"otze, Leutheusser, and
  Yip]{Goetze:1981}
W.~G\"otze, E.~Leutheusser and S.~Yip, \emph{{Phys. Rev. A}}, {1981},
  \textbf{{23}}, {2634--2643}\relax
\mciteBstWouldAddEndPuncttrue
\mciteSetBstMidEndSepPunct{\mcitedefaultmidpunct}
{\mcitedefaultendpunct}{\mcitedefaultseppunct}\relax
\EndOfBibitem
\bibitem[G\"otze \emph{et~al.}({1981})G\"otze, Leutheusser, and
  Yip]{Goetze:1981c}
W.~G\"otze, E.~Leutheusser and S.~Yip, \emph{{Phys. Rev. A}}, {1981},
  \textbf{{24}}, {1008--1015}\relax
\mciteBstWouldAddEndPuncttrue
\mciteSetBstMidEndSepPunct{\mcitedefaultmidpunct}
{\mcitedefaultendpunct}{\mcitedefaultseppunct}\relax
\EndOfBibitem
\bibitem[Ernst and Weijland(1971)]{Ernst:1971a}
M.~H. Ernst and A.~Weijland, \emph{Phys. Lett. A}, 1971, \textbf{34}, 39\relax
\mciteBstWouldAddEndPuncttrue
\mciteSetBstMidEndSepPunct{\mcitedefaultmidpunct}
{\mcitedefaultendpunct}{\mcitedefaultseppunct}\relax
\EndOfBibitem
\bibitem[H{\"o}f\/ling and Franosch(2007)]{Lorentz_LTT:2007}
F.~H{\"o}f\/ling and T.~Franosch, \emph{Phys. Rev. Lett.}, 2007, \textbf{98},
  140601\relax
\mciteBstWouldAddEndPuncttrue
\mciteSetBstMidEndSepPunct{\mcitedefaultmidpunct}
{\mcitedefaultendpunct}{\mcitedefaultseppunct}\relax
\EndOfBibitem
\bibitem[Weijland and van Leeuwen(1968)]{Weijland:1968}
A.~Weijland and J.~M.~J. van Leeuwen, \emph{Physica (Amsterdam)}, 1968,
  \textbf{38}, 35\relax
\mciteBstWouldAddEndPuncttrue
\mciteSetBstMidEndSepPunct{\mcitedefaultmidpunct}
{\mcitedefaultendpunct}{\mcitedefaultseppunct}\relax
\EndOfBibitem
\bibitem[Schnyder \emph{et~al.}({2011})Schnyder, H\"ofling, Franosch, and
  Voigtmann]{Schnyder:2011}
S.~K. Schnyder, F.~H\"ofling, T.~Franosch and {\relax Th}.~Voigtmann, \emph{{J.
  Phys.: Condens. Matter}}, {2011}, \textbf{{23}}, {234121}\relax
\mciteBstWouldAddEndPuncttrue
\mciteSetBstMidEndSepPunct{\mcitedefaultmidpunct}
{\mcitedefaultendpunct}{\mcitedefaultseppunct}\relax
\EndOfBibitem
\bibitem[Kert{\'e}sz and Metzger(1983)]{Kertesz:1983}
J.~Kert{\'e}sz and J.~Metzger, \emph{J.~Phys.~A}, 1983, \textbf{16},
  L735--L739\relax
\mciteBstWouldAddEndPuncttrue
\mciteSetBstMidEndSepPunct{\mcitedefaultmidpunct}
{\mcitedefaultendpunct}{\mcitedefaultseppunct}\relax
\EndOfBibitem
\bibitem[Lorentz(1905)]{Lorentz:1905}
H.~A. Lorentz, \emph{Arch. N{\'e}erl. Sci. Exact Natur.}, 1905, \textbf{10},
  336--370\relax
\mciteBstWouldAddEndPuncttrue
\mciteSetBstMidEndSepPunct{\mcitedefaultmidpunct}
{\mcitedefaultendpunct}{\mcitedefaultseppunct}\relax
\EndOfBibitem
\bibitem[H\"of{}ling and Franosch(2012)]{Anomalous_ROPP:2012}
F.~H\"of{}ling and T.~Franosch, \emph{Anomalous transport in the crowded world
  of biological cells}, 2012, \emph{Rep. Prog. Phys.}, under review\relax
\mciteBstWouldAddEndPuncttrue
\mciteSetBstMidEndSepPunct{\mcitedefaultmidpunct}
{\mcitedefaultendpunct}{\mcitedefaultseppunct}\relax
\EndOfBibitem
\bibitem[Stauffer and Aharony(1994)]{Stauffer:Percolation}
D.~Stauffer and A.~Aharony, \emph{Introduction to Percolation Theory}, Taylor
  \& Francis, London, 2nd edn., 1994\relax
\mciteBstWouldAddEndPuncttrue
\mciteSetBstMidEndSepPunct{\mcitedefaultmidpunct}
{\mcitedefaultendpunct}{\mcitedefaultseppunct}\relax
\EndOfBibitem
\bibitem[Jan and Stauffer(1998)]{Jan:1998}
N.~Jan and D.~Stauffer, \emph{Int. J. Mod. Phys. C}, 1998, \textbf{9},
  341--347\relax
\mciteBstWouldAddEndPuncttrue
\mciteSetBstMidEndSepPunct{\mcitedefaultmidpunct}
{\mcitedefaultendpunct}{\mcitedefaultseppunct}\relax
\EndOfBibitem
\bibitem[ben Avraham and Havlin(2000)]{benAvraham:DiffusionInFractals}
D.~ben Avraham and S.~Havlin, \emph{Diffusion and Reactions in Fractals and
  Disordered Systems}, Cambridge University Press, Cambridge, 2000\relax
\mciteBstWouldAddEndPuncttrue
\mciteSetBstMidEndSepPunct{\mcitedefaultmidpunct}
{\mcitedefaultendpunct}{\mcitedefaultseppunct}\relax
\EndOfBibitem
\bibitem[Machta and Moore(1985)]{Machta:1985}
J.~Machta and S.~M. Moore, \emph{Phys. Rev. A}, 1985, \textbf{32}, 3164\relax
\mciteBstWouldAddEndPuncttrue
\mciteSetBstMidEndSepPunct{\mcitedefaultmidpunct}
{\mcitedefaultendpunct}{\mcitedefaultseppunct}\relax
\EndOfBibitem
\bibitem[Hansen and McDonald(2006)]{Hansen:SimpleLiquids}
J.-P. Hansen and I.~McDonald, \emph{Theory of Simple Liquids}, Academic Press,
  Amsterdam, 3rd edn., 2006\relax
\mciteBstWouldAddEndPuncttrue
\mciteSetBstMidEndSepPunct{\mcitedefaultmidpunct}
{\mcitedefaultendpunct}{\mcitedefaultseppunct}\relax
\EndOfBibitem
\bibitem[H{\"o}f{}ling \emph{et~al.}(2011)H{\"o}f{}ling, Bamberg, and
  Franosch]{FCS_scaling:2011}
F.~H{\"o}f{}ling, K.-U. Bamberg and T.~Franosch, \emph{Soft Matter}, 2011,
  \textbf{7}, 1358--1363\relax
\mciteBstWouldAddEndPuncttrue
\mciteSetBstMidEndSepPunct{\mcitedefaultmidpunct}
{\mcitedefaultendpunct}{\mcitedefaultseppunct}\relax
\EndOfBibitem
\bibitem[Kammerer \emph{et~al.}(2008)Kammerer, H{\"o}f\/ling, and
  Franosch]{Percolation_EPL:2008}
A.~Kammerer, F.~H{\"o}f\/ling and T.~Franosch, \emph{Europhys. Lett. (EPL)},
  2008, \textbf{84}, 66002\relax
\mciteBstWouldAddEndPuncttrue
\mciteSetBstMidEndSepPunct{\mcitedefaultmidpunct}
{\mcitedefaultendpunct}{\mcitedefaultseppunct}\relax
\EndOfBibitem
\bibitem[Franosch and Voigtmann(2002)]{Franosch:2002}
T.~Franosch and {\relax Th}.~Voigtmann, \emph{J. Stat. Phys.}, 2002,
  \textbf{109}, 237\relax
\mciteBstWouldAddEndPuncttrue
\mciteSetBstMidEndSepPunct{\mcitedefaultmidpunct}
{\mcitedefaultendpunct}{\mcitedefaultseppunct}\relax
\EndOfBibitem
\bibitem[Franosch \emph{et~al.}(2010)Franosch, H{\"o}f{}ling, Bauer, and
  Frey]{Lorentz_VACF:2010}
T.~Franosch, F.~H{\"o}f{}ling, T.~Bauer and E.~Frey, \emph{Chem. Phys.}, 2010,
  \textbf{375}, 540--547\relax
\mciteBstWouldAddEndPuncttrue
\mciteSetBstMidEndSepPunct{\mcitedefaultmidpunct}
{\mcitedefaultendpunct}{\mcitedefaultseppunct}\relax
\EndOfBibitem
\bibitem[Spanner(2010)]{Spanner:thesis}
M.~Spanner, \emph{Diploma thesis}, Friedrich-Alexander-Universit\"at
  Erlangen-N\"urnberg, 2010\relax
\mciteBstWouldAddEndPuncttrue
\mciteSetBstMidEndSepPunct{\mcitedefaultmidpunct}
{\mcitedefaultendpunct}{\mcitedefaultseppunct}\relax
\EndOfBibitem
\bibitem[{Colberg} and {H{\"o}fling}(2011)]{Colberg:2011}
P.~H. {Colberg} and F.~{H{\"o}fling}, \emph{Comput. Phys. Commun.}, 2011,
  \textbf{182}, 1120--1129\relax
\mciteBstWouldAddEndPuncttrue
\mciteSetBstMidEndSepPunct{\mcitedefaultmidpunct}
{\mcitedefaultendpunct}{\mcitedefaultseppunct}\relax
\EndOfBibitem
\bibitem[Kert{\'e}sz and Metzger(1984)]{Kertesz:1984}
J.~Kert{\'e}sz and J.~Metzger, \emph{J.~Phys.~A}, 1984, \textbf{17},
  L501--L505\relax
\mciteBstWouldAddEndPuncttrue
\mciteSetBstMidEndSepPunct{\mcitedefaultmidpunct}
{\mcitedefaultendpunct}{\mcitedefaultseppunct}\relax
\EndOfBibitem
\bibitem[Kroy \emph{et~al.}(2004)Kroy, Cates, and Poon]{Kroy:2004}
K.~Kroy, M.~E. Cates and W.~C.~K. Poon, \emph{Phys. Rev. Lett.}, 2004,
  \textbf{92}, 148302\relax
\mciteBstWouldAddEndPuncttrue
\mciteSetBstMidEndSepPunct{\mcitedefaultmidpunct}
{\mcitedefaultendpunct}{\mcitedefaultseppunct}\relax
\EndOfBibitem
\bibitem[Szamel(2007)]{Szamel:2007}
G.~Szamel, \emph{The Journal of Chemical Physics}, 2007, \textbf{127},
  084515\relax
\mciteBstWouldAddEndPuncttrue
\mciteSetBstMidEndSepPunct{\mcitedefaultmidpunct}
{\mcitedefaultendpunct}{\mcitedefaultseppunct}\relax
\EndOfBibitem
\end{mcitethebibliography}
\end{document}